# Geometric assembly of topology in hybrid bosonic bands: magnon–phonon excitations in skyrmion crystals

Doried Ghader*[1] and Bilal Jabakhanji[1]

[1] College of Engineering and Technology, American University of the Middle East, Egaila 54200, Kuwait

*doried.ghader@aum.edu.kw

## Abstract

Hybridization between bosonic excitations can produce topological bands, but their Chern numbers conceal how constituent modes generate the resulting Berry geometry. Here we derive an exact decomposition of Berry curvature in multicomponent bosonic Bogoliubov–de Gennes systems into weighted constituent-sector contributions and a term associated with single-quasiparticle entanglement. The latter couples momentum-space variations of the sector imbalance to a gauge-invariant relative Berry connection and has an equivalent gauge-covariant pseudospin representation. Applying this framework to magnon–phonon excitations in a ferromagnetic skyrmion crystal, we find that magnetoelastic coupling generates topological low-energy hybrid bands despite the trivial topology of the two lowest uncoupled magnon bands and degeneracies among folded phonons. The resulting Chern numbers are assembled through constructive, entanglement-assisted, and entanglement-compensated geometric mechanisms. In particular, bands with identical Chern numbers can exhibit opposite entanglement contributions. Our approach thus resolves nonquantized internal geometric contributions hidden within quantized hybrid-band invariants, without requiring an effective two-level Hamiltonian.

## Introduction

The geometry and topology of collective excitations have become central themes in the study of quantum materials, extending concepts originally developed for electronic bands to charge-neutral bosonic quasiparticles. Magnons, in particular, can acquire nontrivial Berry curvature and Chern numbers, giving rise to phenomena such as transverse thermal transport and chiral edge propagation[1–4]. Beyond isolated magnetic excitations, the coupling of magnons to lattice vibrations produces hybrid quasiparticles, commonly referred to as magnon polarons, whose character combines magnetic and elastic degrees of freedom[5–7]. Experimental evidence includes characteristic magnon-polaron anomalies in spin-Seebeck transport[8] and spectroscopic observations of strong magnon–phonon coupling in layered antiferromagnets[9,10]. Raman studies have also revealed hybridization between phonons and two-magnon excitations in $MnPSe_3$[11]. Such hybridization provides an additional route to engineering bosonic band geometry, since the momentum-dependent coupling between different physical excitations can reconstruct their spectra and redistribute Berry curvature. Understanding this interplay is particularly important in multicomponent systems, where the resulting hybrid eigenmodes possess an internal structure that cannot generally be described by either uncoupled sector alone.

The topological properties of magnon–phonon hybrid excitations have been investigated in a variety of magnetic systems. Early theoretical studies showed that magnon–phonon coupling can induce nontrivial Berry curvature and thermal Hall responses[12–14]. Go *et al.*[15] demonstrated that magnetoelastic coupling can generate topological magnon-polaron bands in two-dimensional ferromagnets, with field-tunable Chern numbers captured by an effective two-level description. Zhang *et al.*[16] subsequently revealed an $SU(3)$ topological structure arising from the coupling of two magnon modes to a phonon mode in antiferromagnets, extending the momentum-space pseudospin perspective beyond conventional two-band models. Further studies have explored Berry-curvature redistribution, topological band reconstruction, and associated thermal responses in honeycomb, kagome, and other magnetic lattices[17–25]. Recent experiments have further demonstrated magnon–phonon hybridization down to the monolayer limit[26], directly observed topological magnon polarons in a multiferroic material[27], and revealed chirality-selective magnon–phonon hybridization in layered antiferromagnets[28]. Collectively, these developments establish magnetoelastic hybridization as an important mechanism for generating and controlling nontrivial bosonic topology. Equally important, effective pseudospin descriptions have emerged as valuable tools for understanding the topology of hybrid modes within reduced few-band models[15,16,29].

Despite these advances, the Chern number of a hybrid band characterizes its integrated topology without specifying how that topology is assembled from the underlying physical sectors. Effective pseudospin descriptions provide valuable geometric insight when the relevant excitations can be represented by a small number of coupled modes. In more general multicomponent bosonic Bogoliubov–de Gennes (BdG) systems, however, the constituent sector components may themselves possess nontrivial momentum-space geometry, while anomalous couplings further enrich the structure of the hybrid eigenvectors. Recent developments in the symplectic quantum geometry of bosonic Bogoliubov quasiparticles provide a broader foundation for characterizing such systems[30]. Separately, relationships between quantum entanglement and geometric phases have been investigated using Schmidt-based descriptions of bipartite states[31]. These developments motivate a general framework for resolving how the Berry geometries of the constituent physical sectors combine with the additional geometry generated by their coherent hybridization to determine the topology of hybrid bosonic bands.

Here, we develop a general sector-resolved formulation of the Berry geometry of hybrid bosonic BdG bands. Starting directly from the metric-normalized eigenvectors, we identify the contributions associated with the normalized constituent-sector modes and an additional contribution governed by their single-quasiparticle entanglement geometry. The resulting exact decomposition expresses the hybrid-band Berry curvature as the sum of weighted sector curvatures and an entanglement term determined by the momentum-space variation of the sector imbalance and a gauge-invariant relative sector connection. We further introduce a gauge-covariant entanglement pseudospin whose solid-angle density provides an equivalent geometric representation of the entanglement contribution. Unlike conventional effective-Hamiltonian pseudospins[15,16,29], this construction does not require a two-level reduction and accommodates multicomponent bosonic BdG systems with anomalous terms. While the total Chern number is quantized, its individual sector and entanglement contributions are generally noninteger, revealing geometric information that is not contained in the topological invariant alone.

We demonstrate the framework through magnon–phonon hybrid excitations in a two-dimensional ferromagnetic skyrmion crystal (SkX). Since their experimental observation[32], SkXs have provided a rich setting for collective spin excitations shaped by noncollinear magnetic textures and emergent gauge fields[33–39]. Their magnon band topology has been extensively investigated, revealing chiral edge states and field-dependent topological properties[40–45]. Experimental studies have further revealed signatures of skyrmion-induced magnon band topology through neutron spectroscopy and thermal Hall transport[46–48]. Here, we consider a Néel-type SkX on a triangular spin lattice, where lattice vibrations couple to magnetic fluctuations through the displacement dependence of the interfacial Dzyaloshinskii–Moriya interaction (DMI). For the parameters considered, the lowest two uncoupled magnon bands are topologically trivial, while the folded phonon manifold contains degeneracies. Upon magnetoelastic coupling, the four lowest hybrid bands acquire nonzero Chern numbers. Their momentum-space sector composition and entanglement geometry can then be resolved through our decomposition. We demonstrate that hybrid-band topology can emerge through distinct constructive, entanglement-assisted, and entanglement-compensated mechanisms. Particularly revealing are bands with identical Chern numbers but opposite roles of the entanglement contribution in assembling their topology. These findings show that the quantized topology of hybrid bosonic excitations can conceal markedly different internal geometric organizations, which are exposed by resolving the contributions of the constituent sectors and their entanglement geometry.

# Results

## Sector decomposition of hybrid bosonic BdG eigenmodes

Hybridization between distinct bosonic sectors can generate bands whose topology need not be attributable to either sector alone. To determine how the internal composition of a hybrid eigenmode enters its band geometry, we consider a general bosonic BdG system whose physical Nambu space separates into two sectors. We label these sectors $\mathrm{m}$ and $\mathrm{p}$, anticipating the magnon and phonon sectors considered below, although the construction does not depend on their microscopic origin.

For an isolated positive-energy band $n$, the bosonic BdG eigenproblem and normalization are

$$\Sigma\mathcal{H}_{\mathrm{BdG}}(\boldsymbol{k})\boldsymbol{\Phi}_{n\boldsymbol{k}} = E_n(\boldsymbol{k})\boldsymbol{\Phi}_{n\boldsymbol{k}}, \qquad \boldsymbol{\Phi}_{n\boldsymbol{k}}^{\dagger}\Sigma\boldsymbol{\Phi}_{n\boldsymbol{k}} = 1, \tag{1}$$

where the metric and eigenvector are resolved according to the physical sectors,

$$\Sigma = \begin{pmatrix} \Sigma_{\mathrm{m}} & 0 \\ 0 & \Sigma_{\mathrm{p}} \end{pmatrix}, \qquad \boldsymbol{\Phi}_{n\boldsymbol{k}} = \begin{pmatrix} \boldsymbol{\Phi}_{n\boldsymbol{k}}^{\mathrm{m}} \\ \boldsymbol{\Phi}_{n\boldsymbol{k}}^{\mathrm{p}} \end{pmatrix}. \tag{2}$$

The corresponding sector weights are defined directly from the components of the interacting eigenvector,

$$w_{\mathrm{m}}^{(n)}(\boldsymbol{k}) = (\boldsymbol{\Phi}_{n\boldsymbol{k}}^{\mathrm{m}})^{\dagger}\Sigma_{\mathrm{m}}\boldsymbol{\Phi}_{n\boldsymbol{k}}^{\mathrm{m}}, \tag{3a}$$

$$w_{\mathrm{p}}^{(n)}(\boldsymbol{k}) = \left(\boldsymbol{\Phi}_{n\boldsymbol{k}}^{\mathrm{p}}\right)^{\dagger}\Sigma_{\mathrm{p}}\boldsymbol{\Phi}_{n\boldsymbol{k}}^{\mathrm{p}}, \tag{3b}$$

$$w_{\mathrm{m}}^{(n)} + w_{\mathrm{p}}^{(n)} = 1. \tag{3c}$$

When both weights are positive, the sector components can be normalized separately as

$$\boldsymbol{m}_n(\boldsymbol{k}) = \frac{\boldsymbol{\Phi}_{n\boldsymbol{k}}^{\mathrm{m}}}{\sqrt{w_{\mathrm{m}}^{(n)}(\boldsymbol{k})}}, \qquad \boldsymbol{p}_n(\boldsymbol{k}) = \frac{\boldsymbol{\Phi}_{n\boldsymbol{k}}^{\mathrm{p}}}{\sqrt{w_{\mathrm{p}}^{(n)}(\boldsymbol{k})}}, \tag{4}$$

with $\boldsymbol{m}_n^{\dagger}\Sigma_{\mathrm{m}}\boldsymbol{m}_n = \boldsymbol{p}_n^{\dagger}\Sigma_{\mathrm{p}}\boldsymbol{p}_n = 1$.

This sector decomposition is illustrated schematically in Fig. 1a. The construction requires only a physically meaningful partition of the BdG space and therefore applies independently of the magnetic ground state, the number of internal degrees of freedom, or the presence of particle-number-nonconserving terms. The complete BdG conventions and sector construction are given in Supplementary Notes 1 and 2.

## Single-quasiparticle sector entanglement

For nonnegative sector weights, the hybrid mode defines a single quasiparticle distributed coherently between the two sectors. Its state can be written in Schmidt form as

$$|\psi_{n\boldsymbol{k}}\rangle = \sqrt{w_{\mathrm{m}}^{(n)}}\,|m_{n\boldsymbol{k}}\rangle \otimes |0_{\mathrm{p}}\rangle + \sqrt{w_{\mathrm{p}}^{(n)}}\,|0_{\mathrm{m}}\rangle \otimes |p_{n\boldsymbol{k}}\rangle, \tag{5}$$

where $|m_{n\boldsymbol{k}}\rangle$ and $|p_{n\boldsymbol{k}}\rangle$ are the one-quasiparticle states associated with the normalized sector modes. An arbitrary relative phase between the two Schmidt components has been absorbed into the phonon-sector state in Eq. (5); we restore it explicitly below when formulating the relative sector connection.

The Schmidt spectrum is therefore $\left\{w_{\mathrm{m}}^{(n)}(\boldsymbol{k}), w_{p}^{(n)}(\boldsymbol{k})\right\}$ and the reduced von Neumann entropy is

$$S_{\mathrm{eig}}^{(n)}(\boldsymbol{k}) = -w_{\mathrm{m}}^{(n)} \ln w_{\mathrm{m}}^{(n)} - w_{\mathrm{p}}^{(n)} \ln w_{\mathrm{p}}^{(n)}. \tag{6}$$

It is useful to introduce the sector imbalance

$$\eta_n(\boldsymbol{k}) = w_{\mathrm{m}}^{(n)}(\boldsymbol{k}) - w_{\mathrm{p}}^{(n)}(\boldsymbol{k}), \tag{7a}$$

so that

$$w_{\mathrm{m}}^{(n)} = \frac{1+\eta_n}{2}, \qquad w_{\mathrm{p}}^{(n)} = \frac{1-\eta_n}{2}. \tag{7b}$$

The limits $\eta_n = +1$ and $\eta_n = -1$ correspond to pure m- and p-sector modes, respectively, whereas $\eta_n = 0$ gives equal sector weights and the maximal entropy $S_{\text{eig}}^{(n)} = \ln 2$.

The entanglement considered here is an eigenvector-level, single-quasiparticle entanglement between physical sectors. The reduced sector density matrices and the corresponding entanglement entropy are derived in Supplementary Note 2.

## Sector and entanglement contributions to Berry curvature

The Berry connection of the hybrid band is calculated using the bosonic metric,

$$\mathcal{A}_\mu^{(n)} = i\boldsymbol{\Phi}_{n\boldsymbol{k}}^\dagger \Sigma \, \partial_\mu \boldsymbol{\Phi}_{n\boldsymbol{k}}, \qquad \partial_\mu \equiv \frac{\partial}{\partial k_\mu}. \tag{8}$$

The normalized sector modes similarly define the sector connections

$$\mathcal{A}_{\mu,\text{m}}^{(n)} = i\boldsymbol{m}_n^\dagger \Sigma_\text{m} \, \partial_\mu \boldsymbol{m}_n, \qquad \mathcal{A}_{\mu,\text{p}}^{(n)} = i\boldsymbol{p}_n^\dagger \Sigma_\text{p} \, \partial_\mu \boldsymbol{p}_n, \tag{9a}$$

and the associated curvatures

$$\Omega_{\mu\nu,s}^{(n)} = \partial_\mu \mathcal{A}_{\nu,s}^{(n)} - \partial_\nu \mathcal{A}_{\mu,s}^{(n)}, \; s \in \{\text{m}, \text{p}\}. \tag{9b}$$

Differentiating the sector-resolved eigenvector gives the central decomposition

$$\Omega_{\mu\nu}^{(n)} = w_\text{m}^{(n)} \Omega_{\mu\nu,\text{m}}^{(n)} + w_\text{p}^{(n)} \Omega_{\mu\nu,\text{p}}^{(n)} + \Omega_{\mu\nu,\text{ent}}^{(n)}, \tag{10}$$

where

$$\Omega_{\mu\nu,\text{ent}}^{(n)} = \frac{1}{2}\left[(\partial_\mu \eta_n)\Lambda_\nu^{(n)} - (\partial_\nu \eta_n)\Lambda_\mu^{(n)}\right]. \tag{11}$$

Here,

$$\Lambda_\mu^{(n)} = \mathcal{A}_{\mu,\text{m}}^{(n)} - \mathcal{A}_{\mu,\text{p}}^{(n)} + \partial_\mu \varphi_n \tag{12}$$

is the gauge-invariant relative sector connection, with $\varphi_n$ denoting an explicitly retained relative phase between the two Schmidt components. In a gauge in which this phase is absorbed into the sector modes, Eq. (12) reduces to the difference between the sector connections.

The first two terms in Eq. (10) describe Berry curvature carried by the normalized sector components of the hybrid eigenmode. They should not generally be identified with the Berry curvatures of isolated uncoupled bands. In particular, each sector component may be a momentum-dependent superposition of several uncoupled modes, including modes belonging to a degenerate manifold. By contrast, the entanglement term is not carried by either sector alone. It originates from the coupling between the momentum-space variation of the Schmidt spectrum, represented by $\nabla_{\boldsymbol{k}}\eta_n$, and the relative sector connection.

The geometric content of the decomposition is summarized in Fig. 1b. The total Berry curvature contains a weighted sector contribution and an entanglement contribution generated jointly by the momentum-space variation of the sector imbalance and the relative sector connection.

In two dimensions,

$$\Omega_{xy,\mathrm{ent}}^{(n)} = \frac{1}{2}\hat{\boldsymbol{z}} \cdot \left[\nabla_{\boldsymbol{k}}\eta_n \times \boldsymbol{\Lambda}^{(n)}\right]. \tag{13}$$

Equation (13) shows that strong sector mixing alone does not generate entanglement Berry curvature. A nonzero contribution requires both a momentum-dependent imbalance and a nontrivial relative sector connection. For example, $\Omega_{xy,\mathrm{ent}}^{(n)}$ vanishes locally when $\nabla_{\boldsymbol{k}}\eta_n$ and $\boldsymbol{\Lambda}^{(n)}$ are parallel, even when both sectors have appreciable weight.

The derivation of Eqs. (10)–(13), including the cancellation of the real weight-derivative terms and the gauge transformations of the relative connection, is presented in Supplementary Note 3.

The Chern number consequently separates as

$$C^{(n)} = C_{\mathrm{m}}^{(n)} + C_{\mathrm{p}}^{(n)} + C_{\mathrm{ent}}^{(n)}, \tag{14a}$$

where

$$C_s^{(n)} = \frac{1}{2\pi}\int_{\mathrm{BZ}} w_s^{(n)}(\boldsymbol{k})\Omega_{xy,s}^{(n)}(\boldsymbol{k})\, d^2k\,,\ s \in \{\mathrm{m},\mathrm{p}\}, \tag{14b}$$

and

$$C_{\mathrm{ent}}^{(n)} = \frac{1}{2\pi}\int_{\mathrm{BZ}} \Omega_{xy,\mathrm{ent}}^{(n)}(\boldsymbol{k})\, d^2k. \tag{14c}$$

Only the total $C^{(n)}$ is required to be quantized. The three terms on the right-hand side of Eq. (14a) are generally noninteger geometric contributions whose sum reconstructs the topological invariant. Their purpose is therefore not to divide the Chern number into independently protected indices, but to identify the internal geometric mechanism by which the hybrid-band topology is assembled.

## Gauge-covariant entanglement pseudospin

The structure of Eq. (13) admits a geometric representation. We define the entanglement pseudospin

$$\boldsymbol{n}_{\mathrm{ent}}^{(n)}(\boldsymbol{k}) = \begin{pmatrix} \sqrt{1-\eta_n^2}\cos\varphi_n \\ \sqrt{1-\eta_n^2}\sin\varphi_n \\ \eta_n \end{pmatrix}, \qquad \left|\boldsymbol{n}_{\mathrm{ent}}^{(n)}\right| = 1. \tag{15}$$

Its vertical component is the sector imbalance, $n_{\mathrm{ent},z}^{(n)} = \eta_n$, whereas the transverse amplitude is $2\sqrt{w_{\mathrm{m}}^{(n)} w_{\mathrm{p}}^{(n)}}$.

The resulting unit-vector geometry is depicted conceptually in Fig. 1c. Pure m- and p-sector states occupy opposite poles of the entanglement sphere, whereas equal sector weights lie on its equator. The transverse orientation records the relative-sector phase geometry.

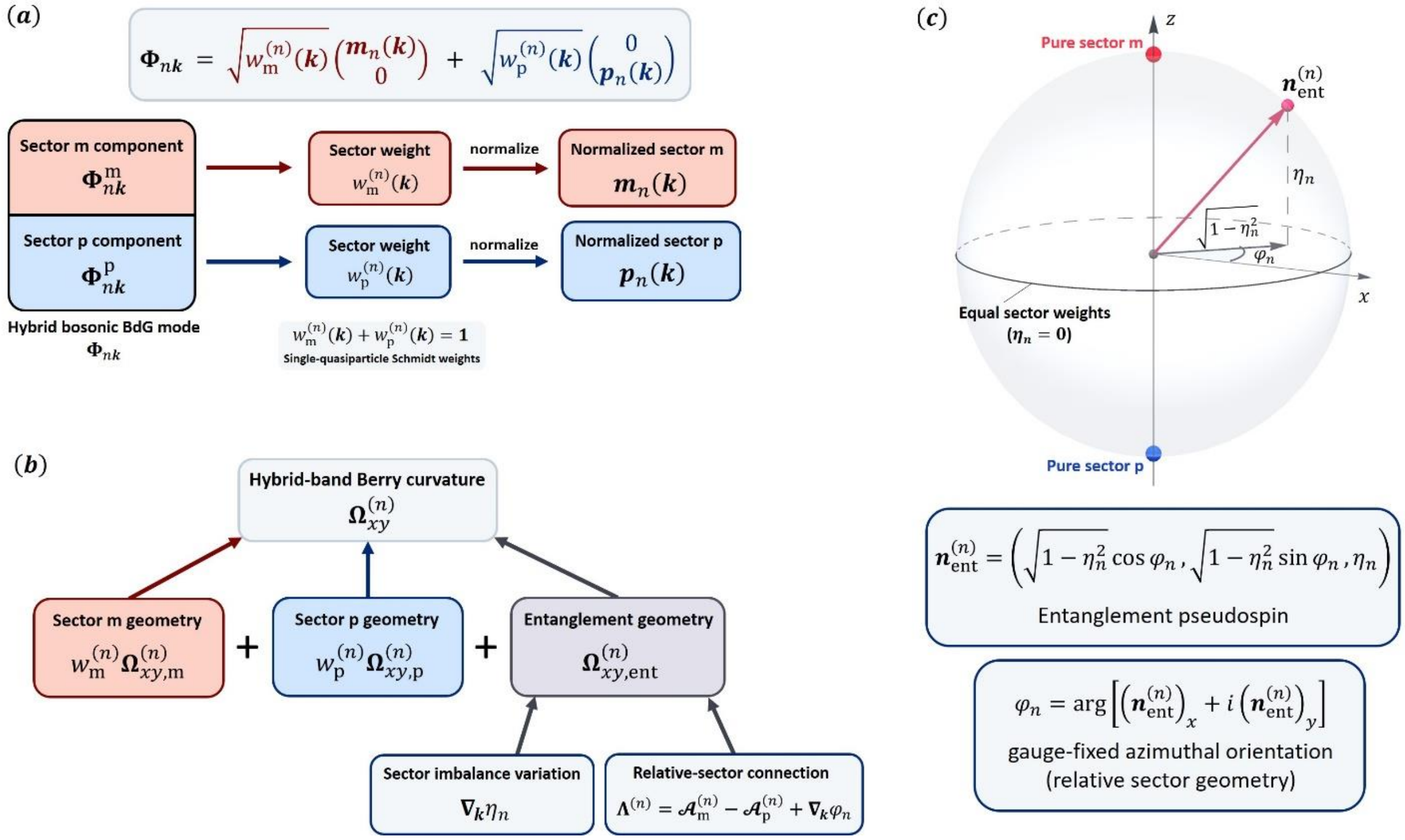

**Fig. 1 | Sector and entanglement geometry of a hybrid bosonic BdG band. a,** Sector-resolved representation of a positive-energy BdG eigenvector in terms of its physical m- and p-sector components, their metric-defined weights $w_m^{(n)}$ and $w_p^{(n)}$, and separately normalized sector modes. When the sector weights are nonnegative, they constitute the Schmidt spectrum of the single hybrid quasiparticle. **b,** Decomposition of the total hybrid-band Berry curvature into weighted sector contributions and an entanglement contribution governed by the momentum-space variation of the sector imbalance $\nabla_{\boldsymbol{k}}\eta_n$ and the relative-sector connection $\boldsymbol{\Lambda}^{(n)}$. **c,** Geometric representation of the entanglement pseudospin $\boldsymbol{n}_{\mathrm{ent}}^{(n)}$ on the unit sphere. The north and south poles correspond to pure m- and p-sector character, respectively, while the equator represents equal sector weights. The longitudinal component is the sector imbalance $\eta_n$, and the transverse magnitude is $\sqrt{1-\eta_n^2}$. The azimuthal angle $\varphi_n$ describes the relative-sector orientation in a chosen gauge, while the associated entanglement Berry curvature is determined by the gauge-covariant pseudospin geometry.

In a general multicomponent BdG problem, the relative connection need not be the gradient of a globally defined scalar phase. The ordinary derivatives of $\boldsymbol{n}_{\mathrm{ent}}^{(n)}$ must therefore be replaced by covariant derivatives $\mathcal{D}_\mu$, whose connection is $\Lambda_\mu^{(n)}$. The entanglement curvature can then be expressed as

$$\Omega_{xy,\mathrm{ent}}^{(n)} = -\frac{1}{2}\boldsymbol{n}_{\mathrm{ent}}^{(n)} \cdot \left(\mathcal{D}_x \boldsymbol{n}_{\mathrm{ent}}^{(n)} \times \mathcal{D}_y \boldsymbol{n}_{\mathrm{ent}}^{(n)}\right). \tag{16}$$

Thus, $\Omega_{xy,\mathrm{ent}}^{(n)}$ is a gauge-invariant solid-angle density constructed from covariant derivatives of the entanglement pseudospin. Equation (16) is exactly equivalent to Eqs. (11) and (13) and therefore provides an alternative route for evaluating both the entanglement Berry curvature and its integrated contribution $C_{\mathrm{ent}}^{(n)}$, provided the covariant derivatives are implemented consistently. In the numerical calculations presented below, we evaluate $\Omega_{xy,\mathrm{ent}}^{(n)}$ using Eq. (11) and obtain $C_{\mathrm{ent}}^{(n)}$ by Brillouin-zone (BZ) integration, while Eq. (16) is used to expose the equivalent solid-angle geometry underlying the entanglement contribution. The equivalence and covariant construction are derived in Supplementary Note 4. Supplementary Note 6 describes the numerical evaluation used for the SkX bands and the construction of their gauge-fixed pseudospin textures.

## Relation to effective-Hamiltonian pseudospin descriptions

The present framework recovers conventional effective-Hamiltonian pseudospin descriptions[15,16,29] as special limits. For a number-conserving two-level Hamiltonian

$$H(\boldsymbol{k}) = d_0(\boldsymbol{k})I + \boldsymbol{d}(\boldsymbol{k}) \cdot \boldsymbol{\sigma}, \tag{17}$$

with momentum-independent physical-sector basis states, the sector Berry curvatures vanish, and the relative-sector connection reduces locally to an ordinary phase gradient. The entanglement pseudospins of the two eigenstates then coincide with the familiar unit vector $\hat{\boldsymbol{d}} = \boldsymbol{d}/|\boldsymbol{d}|$, up to the opposite orientations associated with the two bands, $\boldsymbol{n}_{\mathrm{ent}}^{(\pm)} = \pm\hat{\boldsymbol{d}}$. Consequently, Eq. (16) reduces to the conventional two-band skyrmion-density expression for the Berry curvature, as derived in Supplementary Note 4.

In a general multicomponent bosonic BdG system, however, the physical-sector components of the hybrid eigenvectors may carry nontrivial momentum-space geometry, anomalous couplings can modify their internal structure, and the relative-sector connection need not be represented by a globally defined scalar phase. The entanglement pseudospin is then constructed directly from the full metric-normalized eigenvector, rather than inferred from a reduced effective Hamiltonian. Its gauge-covariant solid-angle density provides the exact entanglement Berry-curvature contribution, which combines with the weighted sector curvatures to reconstruct the total hybrid-band curvature.

This distinction is illustrated in Supplementary Note 4 by revisiting the square-lattice ferromagnetic magnon–phonon model of Go et al.[15]. While its number-conserving limit recovers the conventional $\hat{\boldsymbol{d}}$-texture, the complete bosonic BdG treatment reveals finite sector-geometric contributions that partially compensate the dominant entanglement contribution. Thus, agreement at the level of total Chern numbers does not necessarily imply an equivalent description of the internal geometric assembly of the hybrid bands. The framework likewise complements $SU(3)$ and other reduced multilevel descriptions without requiring the full hybrid Hamiltonian to be represented within a prescribed low-dimensional generator structure.

More generally, the construction is not restricted to a particular magnetic order and can accommodate hybrid excitations above collinear or noncollinear ferromagnetic, antiferromagnetic, and ferrimagnetic ground states. In the remainder of this work, we apply it to a ferromagnetic SkX, where the large magnetic unit cell, folded phonon manifold, and anomalous bosonic couplings provide a stringent multicomponent realization.

## Topological magnon–phonon bands in a ferromagnetic SkX

We now apply the general decomposition to a two-dimensional ferromagnetic SkX on a triangular spin lattice. The microscopic system contains magnetic and elastic degrees of freedom coupled through the displacement dependence of the interfacial DMI,

$$\widehat{\mathcal{H}} = \widehat{\mathcal{H}}_{\mathrm{m}} + \widehat{\mathcal{H}}_{\mathrm{p}} + \widehat{\mathcal{H}}_{\mathrm{mp}}. \tag{18}$$

The magnetic subsystem is described by nearest-neighbor (NN) ferromagnetic exchange, interfacial DMI, and an out-of-plane Zeeman term,

$$\widehat{\mathcal{H}}_{\mathrm{m}} = -J \sum_{\langle i,j \rangle} \widehat{\boldsymbol{\mathcal{S}}}_i \cdot \widehat{\boldsymbol{\mathcal{S}}}_j - d \sum_{\langle i,j \rangle} \widehat{\boldsymbol{d}}_{ij} \cdot \left(\widehat{\boldsymbol{\mathcal{S}}}_i \times \widehat{\boldsymbol{\mathcal{S}}}_j\right) - g\mu_B B \sum_i \hat{\mathcal{S}}_i^z, \tag{19}$$

where $J > 0$, $d > 0$, and the equilibrium interfacial DMI direction is $\widehat{\boldsymbol{d}}_{ij} = \widehat{\boldsymbol{\rho}}_{ij} \times \boldsymbol{e}_z$. Here, $\boldsymbol{\rho}_{ij}$ is the equilibrium NN bond vector, $\widehat{\boldsymbol{\rho}}_{ij} = \boldsymbol{\rho}_{ij}/\left|\boldsymbol{\rho}_{ij}\right| = \boldsymbol{\rho}_{ij}/\ell$, $g$ is the effective Landé factor, $\mu_B$ is the Bohr magneton, and $B$ is the applied magnetic field. We set $g = 1$ throughout when converting between Zeeman energies and magnetic-field values. In the parameter regime considered below, the classical magnetic state is a periodic Néel-type SkX.

The elastic subsystem is modeled by longitudinal NN springs on the same triangular lattice,

$$\widehat{\mathcal{H}}_{\mathrm{p}} = \sum_i \frac{\widehat{\boldsymbol{p}}_i^2}{2M} + \frac{\kappa}{2} \sum_{\langle i,j \rangle} \left[\left(\widehat{\boldsymbol{u}}_i - \widehat{\boldsymbol{u}}_j\right) \cdot \widehat{\boldsymbol{\rho}}_{ij}\right]^2, \tag{20}$$

where $M$ and $\kappa$ are the ionic mass and spring constant, respectively. We retain lattice vibrations polarized along the fixed global $x$-direction, $\widehat{\boldsymbol{u}}_i = \hat{u}_i \boldsymbol{e}_x$ and $\widehat{\boldsymbol{p}}_i = \hat{p}_i \boldsymbol{e}_x$. This restricted polarization preserves the phonon branches involved in the hybridization studied here while keeping the enlarged-supercell BdG problem tractable.

The coupling between the two sectors originates from the dependence of the DMI direction on the instantaneous bond geometry (Fig. 2b). For a small relative displacement $\delta\widehat{\boldsymbol{u}}_{ij} = \widehat{\boldsymbol{u}}_j - \widehat{\boldsymbol{u}}_i$, only the component transverse to the equilibrium bond changes its direction to first order. Because the DMI vector is locked to the bond direction, this generates the magnetoelastic interaction

$$\widehat{\mathcal{H}}_{\mathrm{mp}} = -\frac{d}{\ell}\sum_{\langle i,j\rangle}\left\{\left[\delta\widehat{\boldsymbol{u}}_{ij} - \widehat{\boldsymbol{\rho}}_{ij}\left(\widehat{\boldsymbol{\rho}}_{ij}\cdot\delta\widehat{\boldsymbol{u}}_{ij}\right)\right]\times\boldsymbol{e}_z\right\}\cdot\left(\widehat{\boldsymbol{\mathcal{S}}}_i\times\widehat{\boldsymbol{\mathcal{S}}}_j\right). \quad (21)$$

The periodic magnetic texture defines a triangular SkX superlattice and its corresponding mini-BZ. In the numerical realization considered here, the magnetic unit cell has period $9\ell$ and contains $N_s = 81$ inequivalent microscopic spins, with equivalent boundary sites counted only once (Fig. 2a). Expanding the spin fluctuations about the relaxed noncollinear texture in local Holstein–Primakoff bosons[49], introducing phonon operators on the same $N_s$ - site supercell, and Fourier transforming over the SkX Bravais lattice yields the sector-ordered bosonic BdG Hamiltonian

$$\mathcal{H}_{BdG}(\boldsymbol{k}) = \begin{pmatrix}\mathcal{H}_m(\boldsymbol{k}) & \mathcal{H}_{mp}^{\dagger}(\boldsymbol{k})\\ \mathcal{H}_{mp}(\boldsymbol{k}) & \mathcal{H}_p(\boldsymbol{k})\end{pmatrix}. \quad (22)$$

Both the magnetic noncollinearity and the displacement representation generate anomalous terms, so the full bosonic BdG structure and metric normalization introduced above are retained. The resulting quadratic Hamiltonian is diagonalized within the standard bosonic paraunitary framework[50]. The microscopic model and detailed construction of the magnon, phonon, and magnetoelastic BdG blocks are given in Supplementary Note 5.

We set $J = d = 1\,\mathrm{meV}$ and $S = 1$. Stochastic Landau–Lifshitz–Gilbert simulations using VAMPIRE[51] identify a Néel-type SkX with a characteristic period of $9\ell$ near its minimum stabilizing field, $B_{\min} \simeq 8.2\ \mathrm{T}$. A periodic SkX texture with this characteristic period is then constructed analytically and further relaxed before evaluating the magnon–phonon BdG spectrum, as detailed in Supplementary Note 5.

Figure 2c shows the uncoupled magnon and folded-phonon spectra at $B = 8.2\ \mathrm{T}$. The lowest two magnon bands are topologically trivial, while expressing the elastic problem in the enlarged magnetic unit cell folds the phonon spectrum into the SkX BZ, producing multiple branches and degeneracies. Individual Abelian Chern numbers cannot generally be assigned to the degenerate folded phonon branches in this uncoupled limit.

Including the DMI-induced magnetoelastic interaction reconstructs the low-energy spectrum (Fig. 2d). Crossings between magnon and phonon branches develop avoided crossings, while degeneracies within the folded phonon manifold are also lifted. The resulting gap openings, including the representative features magnified in Fig. 2d, allow the isolated hybrid bands to acquire well-defined Berry curvatures and Chern numbers. Using the gauge-invariant lattice Berry-flux method of Fukui et al.[52], adapted to the bosonic metric as detailed in Supplementary Note 5, we obtain the following Chern numbers for the lowest four hybrid bands

$$\left(C^{(1)}, C^{(2)}, C^{(3)}, C^{(4)}\right) = (-2, 3, 1, 1).$$

The magnetoelastic coupling thus generates nontrivial topology throughout the lowest four hybrid bands, despite the trivial character of the lowest two uncoupled magnon bands. However, the Chern numbers alone do not reveal how the corresponding band geometries assemble from magnon-sector, phonon-sector, and entanglement contributions. The coexistence of distinct integer invariants and differently hybridized eigenmodes within the same microscopic system makes these four bands a particularly instructive realization of the general geometric decomposition. We therefore examine their momentum-space entanglement structure and internal Berry geometry in the following sections.

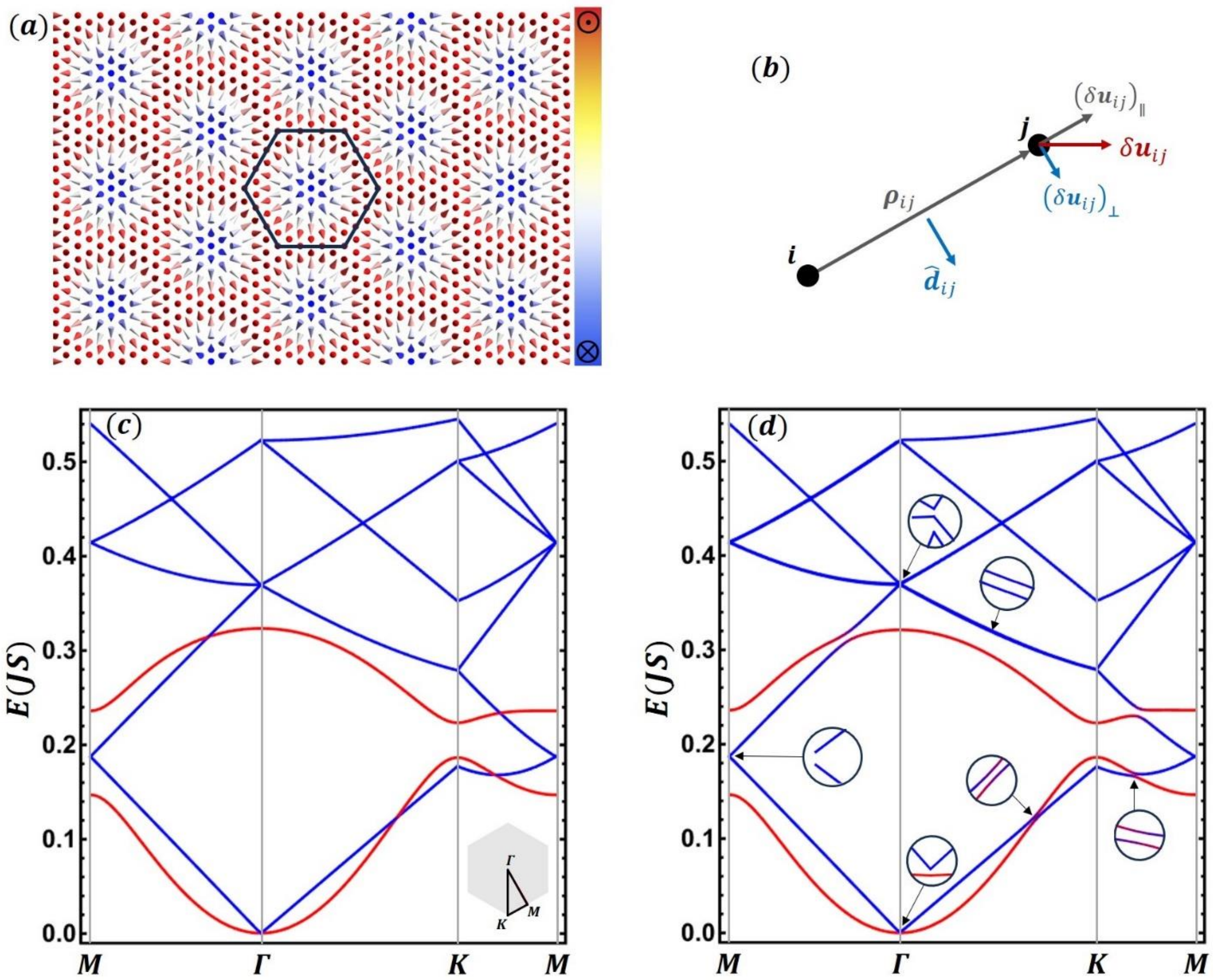

**Fig. 2 | Magnetoelastic activation of topology in a ferromagnetic SkX. a,** Néel-type ferromagnetic SkX on the triangular spin lattice. The outlined magnetic supercell has period $9\ell$ and contains 81 inequivalent spins, with equivalent boundary sites counted only once. Cones indicate the local spin directions, with color representing their out-of-plane components. **b,** Schematic of the DMI-mediated magnetoelastic coupling. For the equilibrium bond connecting sites $i$ and $j$, the interfacial DMI direction is determined by the bond orientation. The relative ionic displacement is decomposed into components parallel and transverse to the equilibrium bond. Only the transverse component modifies the bond and DMI directions to first order. **c,** Uncoupled low-energy magnon (red) and folded-phonon (blue) spectra at $B = 8.2\,T$ along $M - \Gamma - K - M$. The lowest two magnon bands are topologically trivial, while the folded phonon manifold contains degeneracies. The inset identifies the high-symmetry path in the SkX BZ. **d,** Corresponding hybrid spectrum after inclusion of the DMI-induced magnetoelastic coupling. Band colors interpolate between magnon and phonon character. Circular insets magnify representative gaps in the hybrid spectrum. The lowest four hybrid bands carry Chern numbers $(-2, 3, 1, 1)$.

## Momentum-space entanglement and pseudospin textures

The lowest four hybrid bands possess nontrivial Chern numbers, yet their integer invariants do not reveal how the constituent magnon and phonon sectors participate across momentum space. We therefore characterize their internal composition at $B = 8.2\ \mathrm{T}$ through the sector imbalance $\eta_n(\boldsymbol{k})$ and the single-quasiparticle entanglement entropy $S_{\mathrm{eig}}^{(n)}(\boldsymbol{k})$. These quantities provide complementary information. The imbalance distinguishes magnon- and phonon-dominated regions, whereas the entropy measures the degree of sector sharing independently of which sector dominates. The underlying single-quasiparticle Schmidt construction and entanglement entropy are detailed in Supplementary Note 2. Figure 3 focuses on bands 3 and 4, which carry the same Chern number, while the corresponding results for bands 1 and 2 are presented in Supplementary Fig. 2.

Band 3 exhibits an intricate partition between magnon- and phonon-dominated regions (Fig. 3a,c). Extended phonon-rich regions are separated by interconnected magnon-rich channels, producing a network of sharp sector crossovers across the BZ. The entropy is strongly enhanced along these interfaces, where the magnon and phonon weights become comparable. The resulting high-entanglement structures reveal that sector hybridization is distributed through an extended momentum-space network rather than confined to isolated regions.

Band 4 displays a distinctly different organization (Fig. 3b,d). Four prominent magnon-rich regions are separated by a phonon-dominated cross-shaped structure extending through the central BZ. The entropy remains suppressed within the strongly sector-polarized regions and develops pronounced ridges along their interfaces, where the hybrid eigenmode undergoes rapid changes in sector character. Despite their identical Chern numbers, bands 3 and 4 thus exhibit substantially different momentum-space distributions of magnon–phonon entanglement. The approximately twofold symmetry of these maps, also apparent in the Berry-curvature distributions discussed below, arises from retaining only the fixed global $\mathbf{e}_x$ phonon polarization, which selects a preferred in-plane axis despite the hexagonal BZ.

The first two bands provide a complementary contrast (Supplementary Fig. 2). Band 1 is predominantly magnon-like over most of the BZ, with localized phonon-rich regions near the zone boundaries, particularly around the $K$ points. Band 2 exhibits the opposite dominant character, with extensive phonon-rich regions interrupted by localized magnon-rich structures near $K$ and other boundary regions. Their entanglement entropies are correspondingly concentrated near the interfaces separating opposite sector character. These comparatively localized hybridization regions contrast with the more extended and interconnected entanglement landscapes of bands 3 and 4.

The geometric organization of these distributions can be visualized using the entanglement pseudospin introduced in Eq. (15). Figures 3e,f and Supplementary Fig. 2e,f present gauge-fixed pseudospin textures for bands 3 and 4 and bands 1 and 2, respectively. The longitudinal component represents the sector imbalance, while the transverse amplitude is governed by the degree of sector mixing and becomes maximal for equal magnon and phonon weights. The textures consequently exhibit their most pronounced transverse structure near the high-entanglement regions identified in the corresponding entropy maps. Their momentum-space orientations further illustrate the distinct relative-sector organization of the four hybrid bands. The gauge-covariant pseudospin formulation and the numerical construction of the displayed textures are detailed in Supplementary Notes 4 and 6, respectively.

These textures provide a geometric representation of the entanglement structure, but their displayed transverse orientations depend on the prescribed phase-transport convention. Their ordinary winding should therefore not be identified directly with a gauge-invariant topological charge. As established by Eqs. (11) and (16), the entanglement Berry curvature instead depends on the joint momentum-space variation of the sector imbalance and the relative-sector connection, equivalently represented by the gauge-covariant solid-angle density of the entanglement pseudospin. Strong entanglement alone does not determine the magnitude or sign of this curvature.

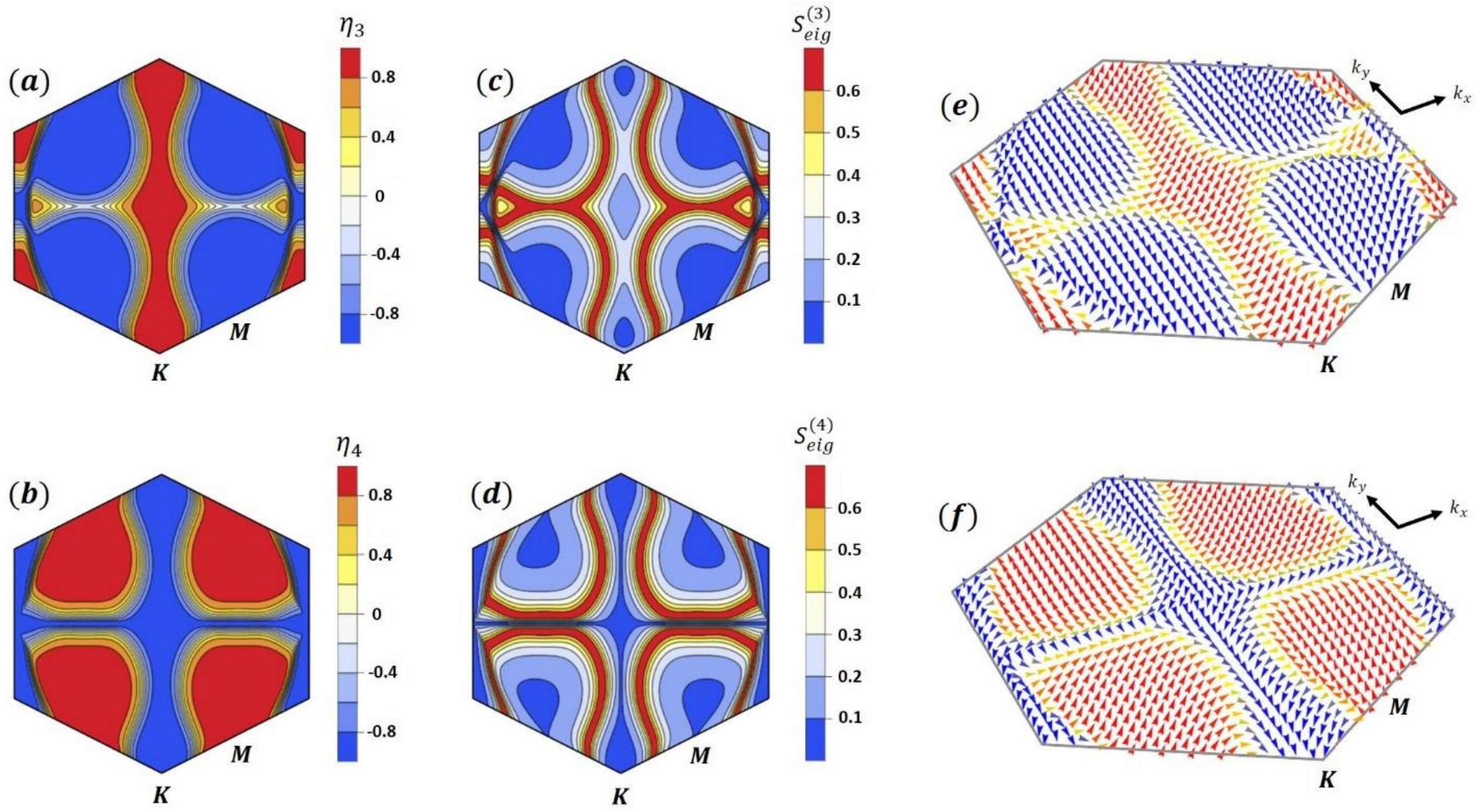

**Fig. 3 | Momentum-space entanglement and pseudospin textures of the topological hybrid bands. a,b,** Sector imbalance $\eta_n(\boldsymbol{k})$ for hybrid bands 3 and 4, respectively, at $B = 8.2\ \mathrm{T}$. Positive (red) and negative (blue) values indicate magnon- and phonon-dominated character, respectively. **c,d,** Corresponding single-quasiparticle entanglement entropy, ranging from zero for pure-sector character to $\ln 2$ for equal sector weights. **e,f,** Gauge-fixed entanglement-pseudospin textures of bands 3 and 4. Three-dimensional vectors are anchored at their corresponding momenta in the SkX BZ, with colors representing the sector imbalance. Their transverse orientations follow the prescribed momentum-space phase-transport convention and visualize the relative-sector organization of the hybrid eigenmodes. The corresponding results for bands 1 and 2 are presented in Supplementary Fig. 2.

The four bands thus exhibit distinct distributions of sector mixing and relative-sector geometry, which are not specified by their Chern numbers. To establish how these differences contribute quantitatively to the topology of the hybrid bands, we next resolve their Berry curvatures into the weighted magnon-sector, phonon-sector, and entanglement contributions.

## Distinct geometric mechanisms of hybrid-band topology assembly

The momentum-space entanglement landscapes in Fig. 3 and Supplementary Fig. 2 reveal how the magnon and phonon sectors are mixed within the four lowest hybrid bands, but they do not determine how this mixing contributes to their Berry geometry. To resolve the internal assembly of the band topology, we decompose the Berry curvature according to Eq. (10) into weighted magnon-sector, phonon-sector, and entanglement contributions. Figure 4 presents the momentum-resolved decomposition of band 4 at $B = 8.2\ \mathrm{T}$, together with the integrated geometric contributions of all four bands. The corresponding momentum-resolved decompositions for bands 1–3 are presented in Supplementary Figs. 3–5, respectively. The numerical procedures are detailed in Supplementary Note 6.

Figures 4a–c show the weighted magnon-sector curvature $w_m^{(4)}\Omega_{xy,m}^{(4)}$, weighted phonon-sector curvature $w_p^{(4)}\Omega_{xy,p}^{(4)}$, and entanglement curvature $\Omega_{xy,\mathrm{ent}}^{(4)}$, respectively. Their distinct momentum-space distributions demonstrate that the constituent geometric contributions are not simply proportional to the full-band curvature. The pointwise sum of these contributions gives the reconstructed curvature $\Omega_{xy,\mathrm{rec}}^{(4)}$ in Fig. 4d, which closely reproduces the independently evaluated full-band curvature $\Omega_{xy}^{(4)}$ in Fig. 4e. This agreement provides a momentum-resolved numerical verification of Eq. (10), while the remaining differences reflect the finite resolution of the independent numerical constructions.

BZ integration reveals three distinct patterns of geometric assembly across the four bands (Fig. 4f). The individual contributions are

$$\left(C_m^{(1)}, C_p^{(1)}, C_{\mathrm{ent}}^{(1)}\right) \simeq (-0.5246, -0.0133, -1.4619),$$

$$\left(C_m^{(2)}, C_p^{(2)}, C_{\mathrm{ent}}^{(2)}\right) \simeq (0.1924, 0.8930, 1.9144),$$

$$\left(C_m^{(3)}, C_p^{(3)}, C_{\mathrm{ent}}^{(3)}\right) \simeq (-0.2200, 0.5743, 0.6427),$$

$$\left(C_m^{(4)}, C_p^{(4)}, C_{\mathrm{ent}}^{(4)}\right) \simeq (0.2587, 1.4507, -0.7073).$$

Their reconstructed sums yield

$$\left(C_{\mathrm{rec}}^{(1)}, C_{\mathrm{rec}}^{(2)}, C_{\mathrm{rec}}^{(3)}, C_{\mathrm{rec}}^{(4)}\right) \simeq (-1.9998, 2.9998, 0.9970, 1.0021),$$

in close agreement with the independently determined integer Chern numbers $(-2, 3, 1, 1)$.

Bands 1 and 2 exhibit constructive, entanglement-dominated assembly. For band 1, all three integrated contributions are negative, with the entanglement term accounting for the largest share of the invariant $C^{(1)} = -2$. The phonon-sector contribution is comparatively negligible, while the magnon-sector contribution reinforces the dominant negative entanglement geometry. For band 2, all three contributions are positive and combine to produce $C^{(2)} = 3$, again with entanglement providing the largest contribution. Despite their opposite Chern-number signs and distinct sector compositions, both bands therefore exhibit sign-coherent geometric assembly dominated by entanglement.

Band 3 realizes a different form of entanglement-assisted assembly. Its positive phonon-sector and entanglement contributions reinforce one another, while the negative magnon-sector term partially compensates their combined contribution. Their balance yields $C^{(3)} = 1$, with entanglement making a substantial positive contribution to the final invariant.

Band 4 instead exhibits entanglement-compensated assembly. Its magnon- and phonon-sector contributions are both positive, with the phonon-sector term alone exceeding the final integer invariant. The entanglement contribution is negative and substantially reduces the combined positive sector geometry, yielding $C^{(4)} = 1$. Thus, bands 3 and 4 share the same quantized Chern number but exhibit opposite roles of entanglement geometry. It assists the positive topology of band 3 while compensating the sector geometry of band 4.

The substantial phonon-sector contributions should not be interpreted as topology inherited from isolated uncoupled phonon bands. The folded phonon manifold contains degeneracies for which individual Abelian Chern numbers are generally undefined. The phonon-sector curvature instead characterizes the normalized phonon component of the interacting hybrid eigenvector, whose momentum-space structure is reconstructed by magnetoelastic coupling. Likewise, the magnon-sector and entanglement contributions are not independently quantized invariants. Their physical significance lies in revealing how different geometric mechanisms combine to produce the same or distinct quantized hybrid-band topologies.

Together, these results show that the four lowest bands of a single SkX system support constructive, entanglement-assisted, and entanglement-compensated mechanisms of geometric assembly. The total Chern number alone cannot distinguish these mechanisms. The sector-resolved decomposition therefore reveals an internal organization of hybrid-band topology that remains hidden in the integer invariants.

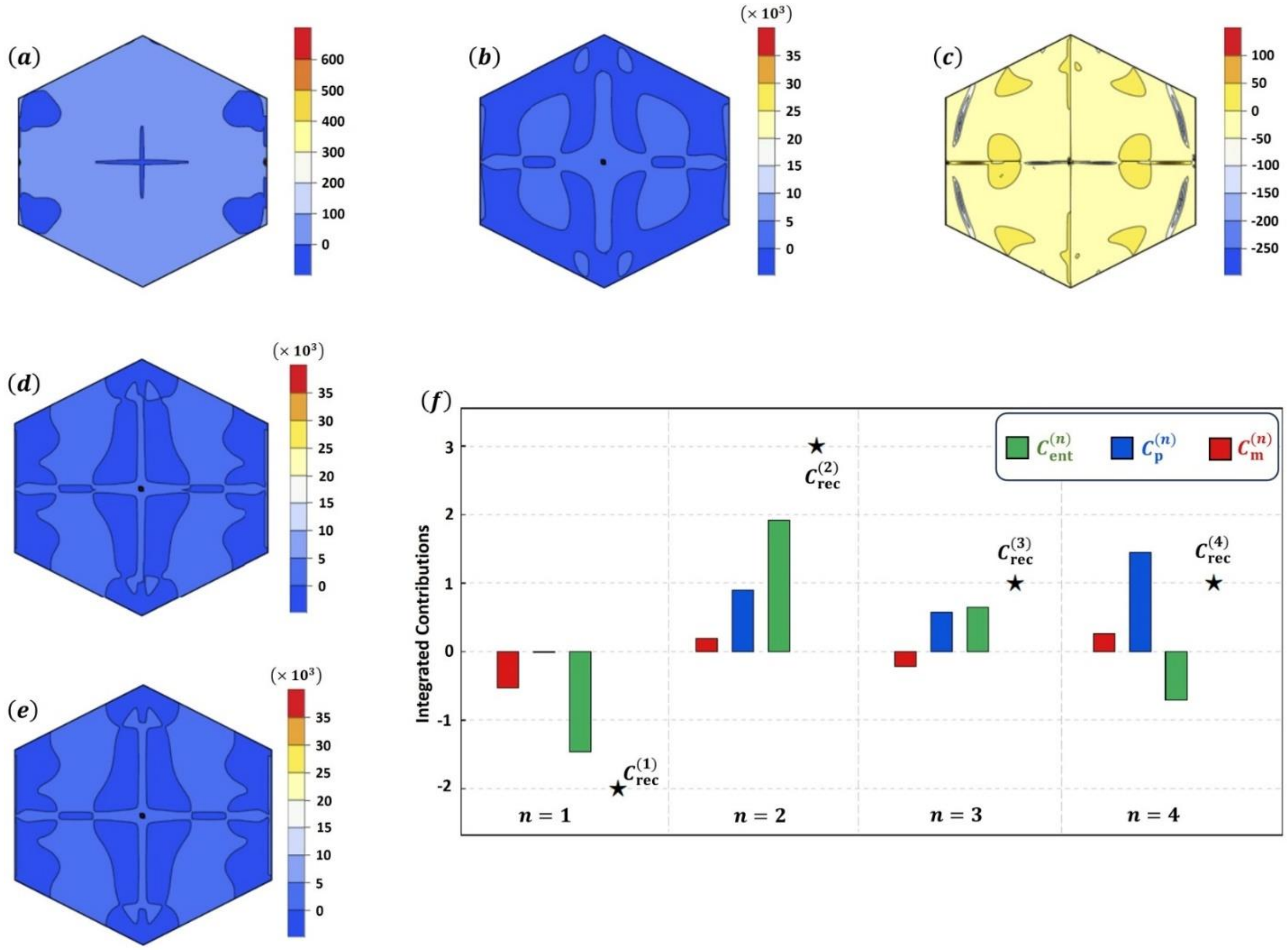

**Fig. 4 | Distinct geometric mechanisms underlying topological magnon–phonon bands. a–c,** Berry-curvature decomposition of hybrid band $n = 4$ at $B = 8.2$ T. **a,** Weighted magnon-sector contribution $w_m^{(4)}\Omega_{xy,m}^{(4)}$. **b,** Weighted phonon-sector contribution $w_p^{(4)}\Omega_{xy,p}^{(4)}$. **c,** Entanglement contribution $\Omega_{xy,\mathrm{ent}}^{(4)}$. **d,** Reconstructed Berry curvature $\Omega_{xy,\mathrm{rec}}^{(4)}$, obtained by summing the three contributions in a–c. **e,** Corresponding full-band Berry curvature $\Omega_{xy}^{(4)}$, evaluated independently using the Fukui-based lattice procedure. The close agreement between d and e demonstrates the momentum-resolved reconstruction of the hybrid-band curvature. **f,** BZ-integrated magnon-sector, phonon-sector, and entanglement contributions for the four lowest hybrid bands. Stars indicate the reconstructed Chern sums, which closely reproduce the independently determined integer invariants $(-2, 3, 1, 1)$. Bands 1 and 2 exhibit constructive, entanglement-dominated assembly, while bands 3 and 4 illustrate entanglement-assisted and entanglement-compensated assembly, respectively. The corresponding momentum-resolved decompositions for bands 1–3 are presented in Supplementary Figs. 3–5.

## Discussion

The present work establishes a geometric framework for understanding how the topology of hybrid bosonic bands emerges from the interplay between their constituent physical sectors. Although the Chern number provides a quantized characterization of an isolated band, it does not reveal how the internal structure of the hybrid eigenmode contributes to that topology. By resolving the bosonic BdG eigenvector into its physical sectors, we obtain an exact decomposition of the Berry curvature into weighted sector contributions and an additional term associated with their single-quasiparticle entanglement geometry. This decomposition provides information beyond the total topological invariant and, importantly, does not rely on a reduced two-level Hamiltonian or the conservation of quasiparticle number.

The application to a ferromagnetic SkX demonstrates that markedly different geometric mechanisms can emerge within the same microscopic system. Depending on the hybrid band, the entanglement contribution can reinforce the geometry carried by the constituent sectors or compensate for it. Particularly revealing is the occurrence of bands with identical Chern numbers but substantially different sector compositions and opposite roles of entanglement geometry in assembling their topology. These findings show that topologically equivalent hybrid bands, as classified by their Chern numbers, can possess distinct internal geometric structures. The individual contributions are not independently quantized invariants, but their decomposition identifies how the quantized total arises from the momentum-dependent organization of the hybrid eigenmodes.

An important implication is that strong magnon–phonon mixing does not necessarily imply a large entanglement contribution to Berry curvature. Although the single-quasiparticle entropy measures the degree of sector sharing, the entanglement curvature depends jointly on the momentum-space variation of the sector imbalance and the relative sector connection. The gauge-covariant entanglement pseudospin provides a geometric interpretation of this relationship through an associated solid-angle density. Unlike the ordinary pseudospin textures of effective two-level models, its transverse orientation generally depends on the chosen gauge representation, while its covariant geometry remains well defined. This distinction makes it possible to separate the magnitude of sector hybridization from the geometric structure through which hybridization contributes to band topology.

The framework also clarifies the relationship between effective pseudospin descriptions and the more general multicomponent geometry of hybrid bosonic excitations. Conventional two-level constructions emerge as special limits in which the sector basis states carry no intrinsic momentum-space curvature. In larger bosonic BdG systems, however, the sector components can develop nontrivial internal geometry, while anomalous couplings and multiple interacting modes prevent a complete description in terms of a simple effective pseudospin texture. The sector-resolved construction accommodates these features directly and is not specific to the DMI-mediated magnetoelastic interaction or to the SkX ground state. It therefore offers a theoretical approach to examine the geometric organization of other coupled bosonic excitations, provided a meaningful physical-sector decomposition and well-defined isolated bands exist.

The present SkX realization employs an idealized periodic magnetic texture and a restricted phonon polarization, while the entanglement characterization concerns the sector-resolved single-quasiparticle BdG eigenmodes. Extensions incorporating additional vibrational polarizations, other coupling mechanisms, and more complex magnetic configurations could reveal further forms of geometric assembly. Experimentally, momentum-resolved inelastic neutron scattering[27] and X-ray scattering could provide complementary information on the magnetic and lattice character of hybrid modes, thereby constraining their sector-weight landscapes through comparison with calculated scattering intensities. Beyond such extensions, the central result is that hybrid-band topology possesses an internal geometric organization that is not captured by its quantized invariant alone. Resolving this organization into constituent-sector and entanglement contributions provides a complementary perspective on topological bosonic excitations and a general framework for understanding how their topology is assembled through hybridization.

## References

1. Katsura, H., Nagaosa, N. & Lee, P. A. Theory of the thermal Hall effect in quantum magnets. *Phys. Rev. Lett.* **104**, 066403 (2010).

2. Shindou, R., Matsumoto, R., Murakami, S. & Ohe, J. Topological chiral magnonic edge mode in a magnonic crystal. *Phys. Rev. B* **87**, 174427 (2013).

3. McClarty, P. A. Topological magnons: a review. *Annu. Rev. Condens. Matter Phys.* **13**, 171–190 (2022).

4. Zhuo, F., Kang, J., Manchon, A. & Cheng, Z. Topological phases in magnonics. *Adv. Phys. Res.* **4**, 2300054 (2025).

5. Flebus, B. *et al.* Magnon-polaron transport in magnetic insulators. *Phys. Rev. B* **95**, 144420 (2017).

6. Streib, S., Vidal-Silva, N., Shen, K. & Bauer, G. E. W. Magnon-phonon interactions in magnetic insulators. *Phys. Rev. B* **99**, 184442 (2019).

7. Bozhko, D. A., Vasyuchka, V. I., Chumak, A. V. & Serga, A. A. Magnon-phonon interactions in magnon spintronics. *Low Temperature Physics* **46**, 383–399 (2020).

8. Kikkawa, T. *et al.* Magnon polarons in the spin Seebeck effect. *Phys. Rev. Lett.* **117**, 207203 (2016).

9. Liu, S. *et al.* Direct observation of magnon-phonon strong coupling in two-dimensional antiferromagnet at high magnetic fields. *Phys. Rev. Lett.* **127**, 097401 (2021).

10. Vaclavkova, D. *et al.* Magnon polarons in the van der Waals antiferromagnet $FePS_3$. *Phys. Rev. B* **104**, 134437 (2021).

11. Mai, T. T. et al. Magnon-phonon hybridization in 2D antiferromagnet $MnPSe_3$. *Sci. Adv.* **7**, eabj3106 (2021).

12. Takahashi, R. & Nagaosa, N. Berry curvature in magnon-phonon hybrid systems. *Phys. Rev. Lett.* **117**, 217205 (2016).

13. Zhang, X., Zhang, Y., Okamoto, S. & Xiao, D. Thermal Hall effect induced by magnon-phonon interactions. *Phys. Rev. Lett.* **123**, 167202 (2019).

14. Park, S., Nagaosa, N. & Yang, B. J. Thermal Hall effect, spin Nernst effect, and spin density induced by a thermal gradient in collinear ferrimagnets from magnon–phonon interaction. *Nano Lett.* **20**, 2741–2746 (2020).

15. Go, G., Kim, S. K. & Lee, K. J. Topological magnon-phonon hybrid excitations in two-dimensional ferromagnets with tunable Chern numbers. *Phys. Rev. Lett.* **123**, 237207 (2019).

16. Zhang, S., Go, G., Lee, K. J. & Kim, S. K. SU(3) topology of magnon-phonon hybridization in 2D antiferromagnets. *Phys. Rev. Lett.* **124**, 147204 (2020).

17. Go, G., Yang, H., Park, J. G. & Kim, S. K. Topological magnon polarons in honeycomb antiferromagnets with spin-flop transition. *Phys. Rev. B* **109**, 184435 (2024).

18. Park, S. & Yang, B. J. Topological magnetoelastic excitations in noncollinear antiferromagnets. *Phys. Rev. B* **99**, 174435 (2019).

19. Thingstad, E., Kamra, A., Brataas, A. & Sudbø, A. Chiral phonon transport induced by topological magnons. *Phys. Rev. Lett.* **122**, 107201 (2019).

20. Sheikhi, B., Kargarian, M. & Langari, A. Hybrid topological magnon-phonon modes in ferromagnetic honeycomb and kagome lattices. *Phys. Rev. B* **104**, 045139 (2021).

21. Huang, H. & Tian, Z. Topological phonon-magnon hybrid excitations in a two-dimensional honeycomb ferromagnet. *Phys. Rev. B* **104**, 064305 (2021).

22. Debnath, S., Bhattacharyya, K. & Basu, S. Topological characterization of magnon-polaron bands and thermal Hall conductivity in a frustrated kagome antiferromagnet. *Phys. Rev. B* **112**, 125404 (2025).

23. Ma, B. & Fiete, G. A. Antiferromagnetic insulators with tunable magnon-polaron Chern numbers induced by in-plane optical phonons. *Phys. Rev. B* **105**, L100402 (2022).

24. Dong, W., Wang, H., Baggioli, M. & Liu, Y. Topological magnetic phases and magnon-phonon hybridization in the presence of strong Dzyaloshinskii-Moriya interaction. *Phys. Rev. B* **113**, 064413 (2026).

25. Mella, J. D., Torres, L. E. F. F. & Troncoso, R. E. Chiral magnon-polaron edge states in Heisenberg-Kitaev magnets. *Phys. Rev. B* **110**, 104433 (2024).

26. Luo, J. *et al.* Evidence for topological magnon–phonon hybridization in a 2D antiferromagnet down to the monolayer limit. *Nano Lett.* **23**, 2023–2030 (2023).

27. Bao, S. *et al.* Direct observation of topological magnon polarons in a multiferroic material. *Nat. Commun.* **14**, 6093 (2023).

28. Cui, J. et al. Chirality selective magnon-phonon hybridization and magnon-induced chiral phonons in a layered zigzag antiferromagnet. *Nat. Commun.* **14**, 3396 (2023).

29. Shen, P. & Kim, S. K. Magnetic field control of topological magnon-polaron bands in two-dimensional ferromagnets. *Phys. Rev. B* **101**, 125111 (2020).

30. Tesfaye, I. & Eckardt, A. Quantum geometry of bosonic Bogoliubov quasiparticles. *Phys. Rev. Res.* **7**, L042052 (2025).

31. Tong, D. M., Sjöqvist, E., Kwek, L. C., Oh, C. H. & Ericsson, M. Relation between geometric phases of entangled bipartite systems and their subsystems. *Phys. Rev. A* **68**, 022106 (2003).

32. Mühlbauer, S. *et al.* Skyrmion lattice in a chiral magnet. *Science* **323**, 915–919 (2009).

33. van Hoogdalem, K. A., Tserkovnyak, Y. & Loss, D. Magnetic texture-induced thermal Hall effects. *Phys. Rev. B* **87**, 024402 (2013).

34. Nagaosa, N. & Tokura, Y. Topological properties and dynamics of magnetic skyrmions. *Nat. Nanotechnol.* **8**, 899–911 (2013).

35. Garst, M., Waizner, J. & Grundler, D. Collective spin excitations of helices and magnetic skyrmions: Review and perspectives of magnonics in non-centrosymmetric magnets. *J. Phys. D: Appl. Phys.* **50**, 293002 (2017).

36. Petrova, O. & Tchernyshyov, O. Spin waves in a skyrmion crystal. *Phys. Rev. B* **84**, 214433 (2011).

37. Mochizuki, M. Spin-wave modes and their intense excitation effects in skyrmion crystals. *Phys. Rev. Lett.* **108**, 017601 (2012).

38. Kim, S. K., Nakata, K., Loss, D. & Tserkovnyak, Y. Tunable magnonic thermal Hall effect in skyrmion crystal phases of ferrimagnets. *Phys. Rev. Lett.* **122**, 057204 (2019).

39. Mook, A., Göbel, B., Henk, J. & Mertig, I. Magnon transport in noncollinear spin textures: Anisotropies and topological magnon Hall effects. *Phys. Rev. B* **95**, 020401 (2017).

40. Roldán-Molina, A., Nunez, A. S. & Fernández-Rossier, J. Topological spin waves in the atomic-scale magnetic skyrmion crystal. *New J. Phys.* **18**, 045015 (2016).

41. Díaz, S. A., Hirosawa, T., Klinovaja, J. & Loss, D. Chiral magnonic edge states in ferromagnetic skyrmion crystals controlled by magnetic fields. *Phys. Rev. Res.* **2**, 013231 (2020).

42. Timofeev, V. E. & Aristov, D. N. Magnon band structure of skyrmion crystals and stereographic projection approach. *Phys. Rev. B* **105**, 024422 (2022).

43. Ghader, D. & Jabakhanji, B. Momentum-space theory for topological magnons in two-dimensional ferromagnetic skyrmion lattices. *Phys. Rev. B* **110**, 184409 (2024).

44. Ghader, D. & Jabakhanji, B. Impact of the honeycomb spin lattice on topological magnons and edge states in ferromagnetic two-dimensional skyrmion crystals. *Phys. Rev. B* **113**, 104430 (2026).

45. Jabakhanji, B. & Ghader, D. Classification of field-driven topological magnon transitions in triangular and honeycomb skyrmion crystals. *J. Phys.: Condens. Matter* **38**, 305803 (2026).

46. Weber, T. *et al.* Topological magnon band structure of emergent Landau levels in a skyrmion lattice. *Science* **375**, 1025–1030 (2022).

47. Akazawa, M. *et al.* Topological thermal Hall effect of magnons in magnetic skyrmion lattice. *Phys. Rev. Res.* **4**, 043085 (2022).

48. Takeda, H. *et al.* Magnon thermal Hall effect via emergent SU(3) flux on the antiferromagnetic skyrmion lattice. *Nat. Commun.* **15**, 566 (2024).

49. Haraldsen, J. T. & Fishman, R. S. Spin rotation technique for non-collinear magnetic systems: application to the generalized Villain model. *J. Phys.: Condens. Matter* **21**, 216001 (2009).

50. Colpa, J. H. P. Diagonalization of the quadratic boson Hamiltonian. *Physica A* **93**, 327–353 (1978).

51. Evans, R. F. L. *et al.* Atomistic spin model simulations of magnetic nanomaterials. *J. Phys.: Condens. Matter* **26**, 103202 (2014).

52. Fukui, T., Hatsugai, Y. & Suzuki, H. Chern numbers in discretized Brillouin zone: efficient method of computing (spin) Hall conductances. *J. Phys. Soc. Jpn.* **74**, 1674–1677 (2005).

## Acknowledgements

Part of the numerical calculations was performed using the Phoenix High-Performance Computing facility at the American University of the Middle East (AUM), Kuwait.